# Dynamic Multiband Microscopy: A Universal Paradigm for Quantitative Nanoscale Metrology

B. N. Slautin,[1,2*] A. Rohi,[1] S. Mathur,[3] A. Ichangi,[3] S. V. Kalinin,[2,4]
D. C. Lupascu,[1] V. V. Shvartsman[1*]

[1] Institute for Materials Science and Center for Nanointegration Duisburg-Essen (CENIDE), University of Duisburg-Essen, Essen, 45141, Germany

[2] Department of Materials Science and Engineering, University of Tennessee, Knoxville, TN 37923, USA

[3] Institute of Inorganic Chemistry, University of Cologne, Cologne, 50939, Germany

[4.] Pacific Northwest National Laboratory, Richland, WA 99354, USA

**Abstract**

Scanning Probe Microscopy (SPM) is the primary tool for exploring nanoscale functionality, yet standard single-frequency operation is fundamentally limited, because the dynamic tip-sample interaction is mathematically underdetermined. While advanced methods such as Dual Amplitude Resonance Tracking (DART) and Band Excitation (BE) address this by tracking resonance, they face critical limitations: DART suffers from feedback instability on complex topographies, while Band Excitation is constrained by severe trade-offs between spectral resolution and acquisition speed. Here, we introduce Dynamic Multiband Microscopy (DMM), a general framework that bridges these gaps by combining multifrequency excitation with continuous frequency sweeping. We implement this within an automated experimental workflow that autonomously identifies and targets measurement points of interest. In combination with quantitative interferometric detection, this approach brings SPM to the fundamental limits of noise and spectral sensitivity. Validated on ferroelectric nanofibers, this platform enables simultaneous, crosstalk-free 3D polarization mapping, establishing a universal framework for autonomous, high-fidelity nanoscale metrology.

---

[*] Authors to whom correspondence should be addressed: bslauti1@utk.edu and vladimir.shvartsman@uni-due.de

**Introduction**

Since its inception, Scanning Probe Microscopy (SPM) has served as the primary gateway to the nanoworld.[1-4] By evolving beyond simple topographic visualization, it has provided a unique capability to probe local physical fields ranging from magnetic and electric potential to mechanical interactions and enabling the exploration of complex physical phenomena with a submicron resolution.[1,5-12] However, the ability to visualize the nanoscale is merely the first step; the transition from qualitative observation to quantitative metrology is the threshold where a technique matures into a rigorous scientific discipline. Just as nanoindentation was transformed from a comparative test into a quantitative science,[13-15] and protein unfolding spectroscopy unlocked the statistical physics of macromolecules,[16-18] SPM must cross the boundary from qualitative imaging to quantitative measurement to meet the demands of modern, data-driven materials discovery.[9]

The heart of quantitative SPM lies in its dynamic detection mechanism. It is well recognized that the conventional dynamic system in single-frequency SPM is mathematically underdetermined, meaning that extracting quantitative mechanical or electromechanical parameters cannot be determined from a single frequency response.[19,20] This realization drove the development of two diverging methodologies to utilize the contact resonance: Dual Amplitude Resonance Tracking (DART) [21] and Band Excitation (BE). [22,23]

In DART, the cantilever is excited at two frequencies located on the flanks of the resonance peak, and an active feedback loop adjusts the drive frequency to minimize the amplitude difference between the channels (Figure 1a). DART became the industry standard for its speed and ease of integration. However, its feedback stability is heavily dependent on the local resonance shape, making it susceptible to artifacts on samples with large variations in topography or properties.[24,25]

Band Excitation, by contrast, employs simultaneous excitation over a finite frequency band around the resonance, eliminating the need for active resonance tracking and enabling higher-fidelity reconstruction of the frequency response (Figure 1b).[22] However, distributing the excitation power across multiple frequencies introduces a trade-off between spectral resolution, signal-to-noise ratio, and acquisition speed. Moreover, careful selection of the excitation bandwidth and frequency grid is required to ensure that resonance shifts remain within the excited range.

The alternative solution may be a continuous resonance sweep, in which the entire resonance curve is acquired frequency by frequency, providing the highest possible fidelity of the electromechanical response (Figure 1c).[26] This method also referred in different sources as R-PFM or "resonance tracking" has been initially proposed by S. Jesse et. al in 2006[27] and has been successfully implemented for the characterization of weak functional responses including, for instance, piesoresponse[27] and local electrostriction.[28] The increase in spectral resolution comes at the cost of longer measurement times and, consequently, reduced spatial resolution (due to the limited number of points per "scan" needed to shorten acquisition time), making this method less suitable for visualization purposes. However, resonance tracking is still applied for statistical characterization, where spatial distributions are not required, including the piezoelectric characterization of nanocrystals[29] and fibers.[30,31]

Summarizing, a persistent gap remains between fast, qualitative imaging and slow, quantitative spectroscopic measurements. This trade-off is particularly detrimental for topographically non-uniform materials, where complex geometries induce strong topographic crosstalk that obscures the weak electromechanical response.

In this work, we introduce Dynamic Multiband Microscopy (DMM) as a general framework to bridge this gap. DMM combines the stability of multifrequency excitation with the spectral resolution of continuous frequency sweeping. By performing a parallel sweep of the frequency domain, this method resolves the "underdetermined" nature of the tip-sample interaction without the stability issues of feedback loops or the spectral limitations of traditional band excitation. We demonstrate the efficacy of DMM on ferroelectric $K_{0.49}Na_{0.51}NbO_3$ (KNN) nanofibers, showing that the technique effectively suppresses topographic crosstalk and separates mechanical and electromechanical contributions. Furthermore, the implementation of a multi-frequency, multi-resonance configuration enables the simultaneous detection of vertical and lateral responses, providing a robust pathway toward quantitative, three-dimensional polarization mapping.[16]

**Results and Discussion**

Dynamic Multiband Microscopy combines multifrequency excitation with continuous frequency sweeping to enable fast and quantitative dynamic SPM measurements. In DMM, the system is excited simultaneously at multiple frequencies, analogous to Band Excitation, while the excitation window is continuously swept across a defined frequency range. This implementation can be realized either using a multifrequency waveform or by employing multiple signal generators, each applying a voltage at a distinct frequency offset within the sweep window (Figure 1d).

This combined excitation–sweep strategy offers several key advantages. Parallel multifrequency excitation substantially reduces sweep time, enabling high-spatial-resolution measurements without sacrificing acquisition efficiency. Simultaneously, continuous sweeping over an extended frequency range ensures robust tracking of resonance shifts in samples with strong mechanical or electromechanical heterogeneity, while preserving high frequency resolution for accurate resonance fitting and reliable extraction of true surface deformation amplitudes.

A multi-generator implementation enables simultaneous sweeping of multiple resonances, including vertical and lateral modes (multi-frequency multi-resonance, MF-MR), with each response recorded independently (Figure 1e). This configuration eliminates misalignment between vertical and lateral signals and enables reliable determination of polarization direction in vector PFM. [32] Similar principles have been explored in early dual-resonance excitation schemes, where multiple cantilever modes were simultaneously driven and analyzed.[33]

The acquisition time of DMM nevertheless remains higher than that of rapid BE or DART imaging, reflecting an inherent trade-off between measurement fidelity and throughput. Consequently, the effective use of DMM benefits from adaptive, automated SPM strategies that selectively target points or regions of interest rather than uniform raster scanning.[34-36] This consideration makes DMM particularly well suited for the quantitative characterization of low-dimensional materials, such as nanoparticles, nanofibers, or nanorods, where only a limited number of spatial locations carry meaningful functional information. In such systems, transitioning from continuous raster imaging to point-to-point or hopping-mode measurements enables efficient deployment of DMM while reducing topography-induced crosstalk and probe–sample perturbations that can obscure weak electromechanical signals. In practice, this targeted approach relies on prior non-contact topographic imaging to identify relevant objects and guide the selection of measurement locations.

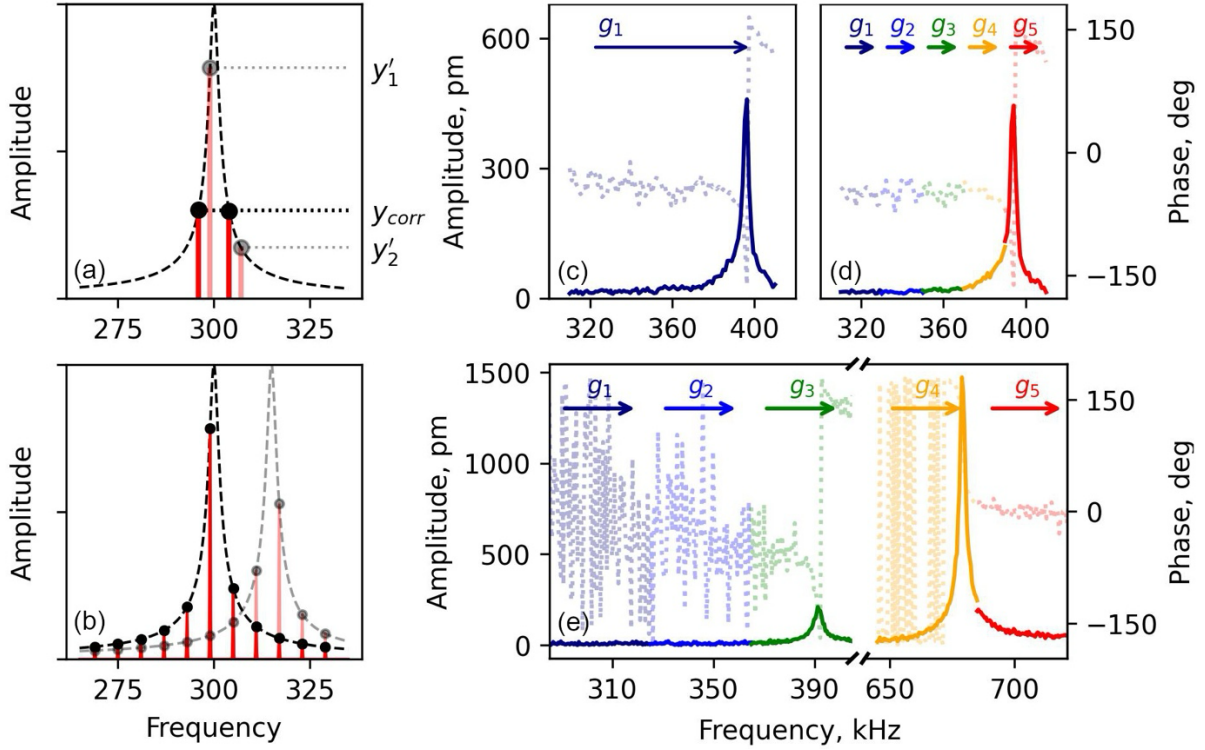

**Figure 1.** (a) DART mode, where black dots represent amplitudes measured at two frequencies located on the wings of the resonance. Semi-transparent points and lines indicate the measured signals before PID-based correction of the driving frequency. (b) Band Excitation mode, where the cantilever response is recorded simultaneously over a band of frequencies spanning the resonance, without active feedback loops. During scanning, shifts in the resonance frequency lead to a redistribution of the response amplitude among the excited frequencies. (c) Single-frequency single-resonance (SF-SR) mode (also referred as R-PFM, resonance tracking); (d) Dynamic Multiband Microscopy in multi-frequency single-resonance (MF-SR) mode and (e) in multi-frequency multi-resonance (MF-MR) mode. Individual signal generators are indicated by different colors and labels ($g_x$). In (e), the first resonance set ($g_1 - g_3$) corresponds to the vertical response, while the second set ($g_4 - g_5$) corresponds to the lateral response.

We applied the multi-frequency multi-resonance (MF-MR) approach to assess its capability for quantitative characterization of low-dimensional piezoelectric objects, using $K_{0.49}Na_{0.51}NbO_3$ (KNN) fibers as a representative test system. As an initial approximation, the fibers were treated as quasi-one-dimensional structures, such that measurements were performed along their longitudinal axis, with densely sampled points forming a linear sequence. This assumption implicitly neglects possible polarization rotation across the fiber width and restricts domain variations to the axial direction. Within this framework, the MF-MR methodology enables a substantial reduction in the number of measurement points by probing a single representative location at each cross-section along the fiber length.

Under these conditions, the reconstructed surface displacement amplitudes and local phase maps resolve the domain structure along the fiber axis (Figure 2). The bimodal phase distributions observed in both vertical and lateral channels correspond to opposite polarization projections along the respective directions, demonstrating the ability of MF-MR to simultaneously capture vector electromechanical information. In parallel, the locally extracted resonance frequencies reveal variations in contact stiffness, with higher resonance frequencies observed for fibers located beneath others and near fiber intersections, indicating increased mechanical constraints in these regions. While the Q factor is rarely analyzed in dynamic SPM due to its limited interpretability, DMM measurements enable its reliable extraction, providing additional insight into local energy dissipation and mechanical boundary conditions

The MF-MR measurements also expose the limitations of the one-dimensional approximation. For an ideal one-dimensional piezoelectric fiber, the PFM amplitude would be expected to vary systematically with probe–fiber orientation. The absence of such dependence in the experimental data indicates that the electromechanical response is governed by local effects at sub-fiber length scales, including nanoscale domain configurations and microstructural heterogeneity, rather than by the global fiber geometry. This observation highlights the necessity of extending beyond purely axial measurements and motivates grid-based MF-MR mapping within individual fibers to resolve the coupled evolution of domain structure and local microstructure.

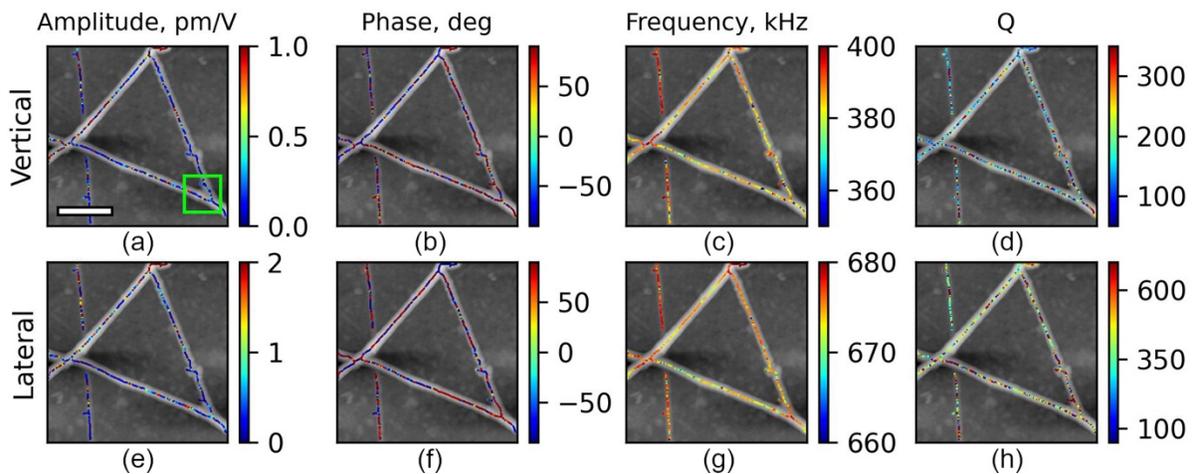

**Figure 2.** MF–MR PFM characterization of vertical and lateral piezoresponse in KNN nanofibers in 1D sample policy modes. (a,e) Amplitude, (b,f) phase, (c,g) resonance frequency, and (d,h)

quality factor for the vertical and lateral responses, respectively. The total experiment duration was below 1 h. Scalebar: 1.5 μm.

To evaluate the capability of the DMM approach for resolving local electromechanical complexity beyond one-dimensional assumptions, we performed grid-based measurements within a single, selected local region of an individual KNN fiber (Figure 3). A representative area of 1x1 μm was chosen to probe spatial variations across and along the fiber while maintaining high spectral fidelity. This targeted strategy limits the total number of measurement points while enabling multidimensional characterization within a physically meaningful region.

Within this local grid, the PFM amplitude and phase maps acquired in both vertical and lateral channels reveal a pronounced and highly heterogeneous domain structure (Figure 3c-j). Alternating contrast in the phase images indicates the presence of domains with opposite polarization components both along and across the fiber axis. While precise reconstruction of the full polarization vector distribution requires careful alignment and overlapping of scans acquired at different probe–fiber orientations,[32] a simplified "reduced" vector PFM can provide its approximate estimation. By jointly analyzing the fully compatible vertical and lateral electromechanical responses, we reconstruct the local projection of the polarization vector onto the *xz* plane, providing an effective proxy for 3D vector PFM (Figure 3a,b).

The reconstructed polarization maps demonstrate that the local polarization direction varies substantially within the selected region and is not constrained to the fiber axis. Instead, the electromechanical response is governed by local factors, including microstructural heterogeneity and mechanical boundary conditions. Consistent with this interpretation, the average lateral piezoresponse ranges from 1 to 2 pm/V and shows no systematic dependence on fiber orientation (Figure 3g), while the vertical response remains below 1 pm/V (Figure 3c). This asymmetry may reflect substrate-induced mechanical clamping, which suppresses out-of-plane deformation while leaving shear deformation comparatively less constrained.

The simultaneously extracted resonance frequency and quality factor maps provide complementary information on local mechanical and dissipative behavior (Figure 3e,f,i,j). Elevated resonance frequencies are observed at grain junctions indicating locally increased stiffness and stronger tip–sample coupling. Mechanical variations that would typically introduce strong topographic crosstalk are effectively decoupled from the electromechanical response in the DMM MF-MR framework. Regions within the same polarization domain can nevertheless exhibit distinct quality factors, underscoring the separation of mechanical and electromechanical contributions achieved by the method.

Together, these results demonstrate that even within a nominally one-dimensional KNN fiber, the electromechanical response is strongly modulated by local microstructure and mechanical constraints. DMM MF-MR measurements therefore enable reconstruction of cross-talk free polarization projections and quantitative electromechanical characterization without resorting to exhaustive scanning of the entire structure.

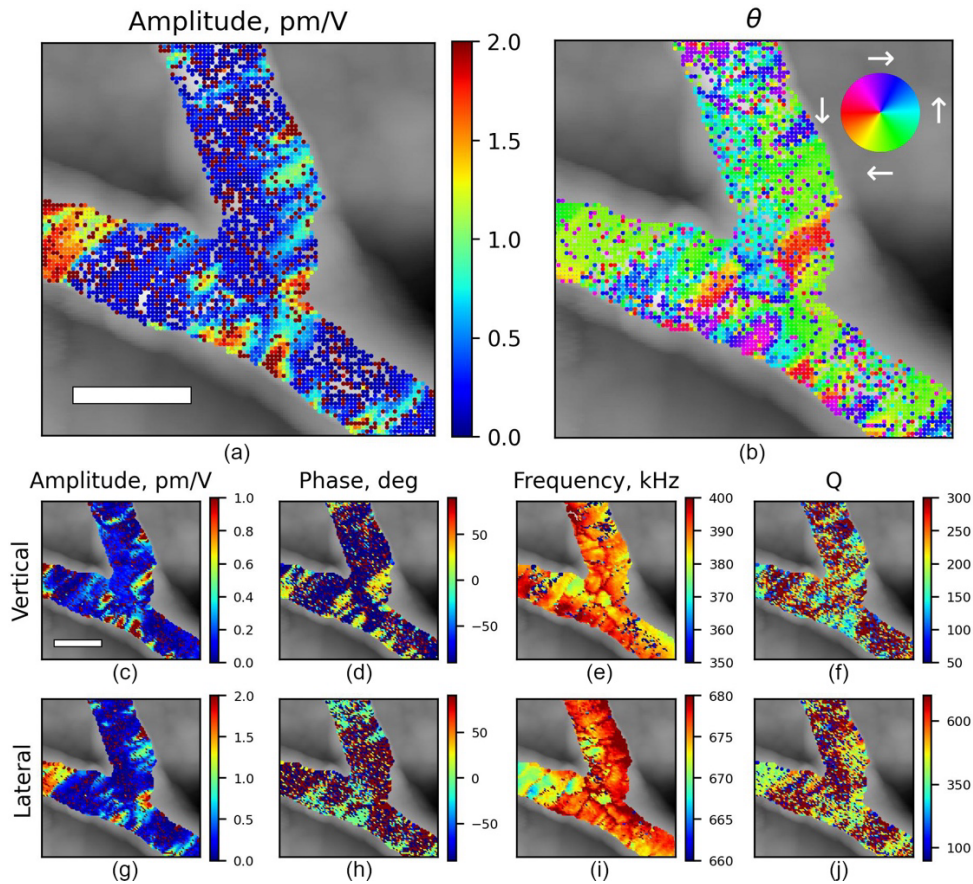

**Figure 3.** (a) Total polarization amplitude and (b) polarization angle projected onto the *xz* plane (reduced 3D PFM) in KNN nanofibers. (c–j) Multifrequency–multiresonance (MF–MR) PFM maps of the vertical and lateral electromechanical response, including amplitude, phase, and resonance-frequency distributions acquired at the same sample region. The total acquisition time for each map was approximately 1–2 h. Scale bar: 300 nm.

Simultaneous measurement of the in-plane and out-of-plane phase signals enables identification of the polarization orientation within one of four spatial quadrants (up–left, up–right,

down–left, down–right), but does not resolve variations along the longitudinal axis. Nevertheless, this rapid estimation provides a useful proxy for capturing the overall characteristics of the domain structure.

Combined analysis of the vertical and lateral phase responses indicates that KNN fibers exhibit a three-dimensional domain configuration while largely preserving a tendency toward full–fiber-width domains (Figure 4). Notably, domains with polarization oriented in the "down-right" quadrant were largely absent and appeared only as small interfacial regions between neighboring domains. Although the current dataset does not allow for a definitive conclusion, this observation may suggest the existence of a preferred global polarization arrangement within the fiber architecture.

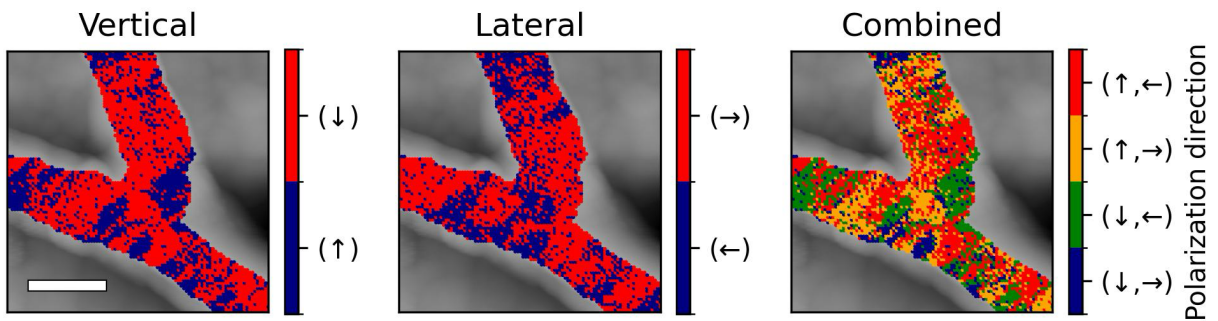

**Figure 4.** Approximate domain reconstruction from binarized phase maps of vertical and lateral piezoresponse. Combined representation of the resulting approximate domain pattern. Scale bar: 300 nm.

**Summary**

In this work, we introduced Dynamic Multiband Microscopy as a versatile dynamic detection approach that restores quantitative determinacy to scanning probe microscopy. By combining multifrequency excitation with continuous frequency sweeping, DMM addresses the trade-off between stability, spectral resolution, and acquisition speed that has constrained resonance-based SPM methods for nearly two decades. Unlike feedback-driven approaches or fixed-band excitation schemes, DMM enables robust reconstruction of resonance parameters while remaining resilient to strong topographic and mechanical heterogeneity.

The implementation of a multi-frequency, multi-resonance (MF–MR) configuration represents a key advance, allowing the simultaneous and inherently aligned detection of vertical and lateral electromechanical responses. This capability enables crosstalk-free, three-dimensional polarization mapping within a single measurement framework and provides direct access to local electromechanical, mechanical, and dissipative properties. Importantly, the method decouples electromechanical contrast from variations in contact stiffness and damping, a critical requirement for quantitative interpretation on structurally complex or low-dimensional systems.

Application of DMM to $K_{0.49}Na_{0.51}NbO_3$ nanofibers reveals that their electromechanical response is governed by local microstructural heterogeneity and mechanical boundary conditions rather than by simple one-dimensional polarization alignment. The observed dominance of in-plane polarization components, suppression of out-of-plane response, and multidimensional domain configurations underscore the necessity of quantitative, resonance-resolved measurements for reliable interpretation of nanoscale functional behavior. These results further demonstrate that DMM enables meaningful polarization reconstruction without exhaustive raster scanning, making it particularly well suited for sparse, low-dimensional materials.

Beyond the specific case of ferroelectric nanofibers, DMM provides a general, automation-ready framework that can be readily implemented on commercial SPM platforms. Its compatibility with targeted measurements, adaptive sampling, and autonomous workflows positions DMM as a foundational tool for high-throughput and AI-driven nanoscale metrology. By unifying quantitative rigor with experimental practicality, Dynamic Multiband Microscopy establishes a new standard for dynamic SPM and opens a pathway toward reliable, data-centric characterization of functional nanostructures.

**Experimental**

All experiments were performed using an MFP-3D atomic force microscope (Asylum Research, Oxford Instruments). Multi75-EG conductive probes (BudgetSensors) were employed for all measurements. Prior to each experiment, the probes were individually characterized to define individual stiffness and Deflection-Heights's translation coefficient for the quantitative characterization. For multifrequency excitation and signal demodulation, a HF2LI lock-in amplifier (Zurich Instruments) was utilized, providing generation and detection of multiple excitation frequencies within the resonance band.

The microscope operation and data acquisition were automated using a Python script developed with the AESPM library, enabling precise control of probe positioning, waveform generation, and synchronized signal acquisition. For the identification of individual nanofibers, we developed a Python script that operates through a morphological filtering, skeletonization and binarization steps. Information about the implementation is provided in the Supplementary Material.

The experimental validation was performed on potassium sodium niobate nanofibers with a multigrain structure, serving as a benchmark system for low-dimensional ferroelectric materials. Pure KNN fibers with the nominal composition $K_{0.49}Na_{0.51}NbO_3$ were synthesized via the sol–gel electrospinning method. Metal alkoxide precursors of potassium, sodium, and niobium were dissolved in isopropanol with glacial acetic acid serving as a stabilizing agent. Polyvinylpyrrolidone (PVP) was added as a polymeric binder to form a homogeneous spinning solution, which was electrospun under an applied electric field of approximately 1.5 kV cm$^{-1}$. The as-spun fibers were collected on a rotating drum and subsequently calcined in air at 300 °C and 700 °C to remove organic components and crystallize the perovskite KNN phase. The KNN fibers were collected on top of a conductive substrate prepared by sputtering 80 nm of Ti, followed by a 100 nm Pt layer on top of silicon substrates. The detailed description of the synthesis process can be found in the literature. [6]

**Data availability**

All raw experimental data and the full analysis pipeline can be accessed through the GitHub repository: https://github.com/Slautin/2025_Dynamic-Multifrequency-Spectroscopy_fibers.

**Author contribution**

**BNS:** Conceptualization; Software; Data curation; Writing – original draft. **AR:** Conceptualization; Software. **SM:** Resources; **AI:** Resources; **SVK:** Supervision; Writing – review & editing. **DCL:** Writing – review & editing. **VVS:** Conceptualization; Supervision; Writing – review & editing.

**Acknowledgments**

The authors AI and SM acknowledge the financial support from the European Union's Horizon 2020 research and innovation program (ITN ENHANCE) under the Marie Sklodowska-Curie grant (ID: 722496), as well as the excellent infrastructure provided by the University of Cologne. This effort (analysis of measured data) was based upon work supported by the U.S. Department of Energy (DOE), Office of Science, Basic Energy Sciences (BES), Materials Sciences and Engineering Division (B.N.S., S.V.K.).

# Supplementary Material

Dynamic Multiband Microscopy: A Universal Paradigm for Quantitative Nanoscale Metrology

B. N. Slautin, A. Rohi, S. Mathur, A. Ichangi, S. V. Kalinin,
D. C. Lupascu, V. V. Shvartsman

1. Identification of individual nanofibers.

The identification of individual fibers has been realized through morphological filtering, skeletonization and binarization steps. The workflow detects elongated fiber structures, extracts their midlines, and samples points along them at a controlled spacing. The full algorithmic procedure is summarized below.

INPUT: *image, point_distance*
OUTPUT: *midline_points_in_micrometers*

1. **Load** *image*
   # Extract the selected imaging channel from the SPM dataset.
2. **Normalize** *image* **to 0–255**
   # Standardize intensity range for robust filtering.
3. **Gaussian bluring** → *image_blur*
   # Reduce high-frequency noise that may interfere with ridge detection.
4. **Apply Meijering ridge filter** → *ridge_map*
   # Enhance elongated, fiber-like structures (ridge detection).
5. **Otsu thresholding** *ridge_map* → *binary_ridges*
   # Convert the enhanced image into a binary mask of fiber regions.
7. **Remove small connected objects from** *binary_ridges*
   # Eliminate noise and non-fiber artifacts.
8. **Skeletonize** *binary_ridges* → *skeleton*
   # Reduce each fiber to a one-pixel-wide centerline (midline).
9. **Label connected components in** *skeleton* → *labeled_midlines*
   # Identify individual fiber midlines as separate objects.
10. **Uniform sampling with** *point_distance*
11. **Convert pixel spacing into physical micrometer spacing**
12. **Return** *physical_points*
    # Final set of uniformly spaced midline sampling locations in micrometers.